Raphaël LAGUIONIE*
Matthieu RAUCH*
Jean-Yves HASCOET*

# TOOLPATHS PROGRAMMING IN AN INTELLIGENT STEP-NC MANUFACTURING CONTEXT

The current language for CNC programming is G-code which dates from the beginning of the eighties with the norm ISO 6983. With the new technologies, G-code becomes obsolete. It presents drawbacks that create a rupture in the numerical chain at the manufacturing step. A new standard, STEP-NC, aims to overtake these lacks. A STEP-NC file includes all the information for manufacturing, as geometry description of the entities, workplan, machining strategies, tools, etc. For rough pocket milling, the ISO norms propose different kind of classical strategies as bidirectional, parallel or spiral contour, etc. This paper describes a new way of toolpath programming by the repetition of a pattern all along a guide curve. It presents several advantages as building fastness and easiness. The integration of pattern strategies in STEP-NC standard is an other step for the development of these strategies but also for the enrichment of STEP-NC possibilities. A complete STEP-NC numerical chain was built, integrating these pattern strategies. The implementation of this approach of building pattern strategies was made by the development of tools for the complete manufacturing cycle, from the CAD file to the machined part. Several application cases were experimented on machine tool to validate this approach and the efficiency of the developed tools.

## 1. INTRODUCTION

The current research on programming machine tools reaches the goal to improve productivity and production possibilities through the development of new solutions to make data flow all along the numerical chain more simple, to generate optimised manufacturing toolpaths, to develop new machine structures, etc. However, the current language used for programming machine tools is G-code, standardized at the beginning of the Eighties with norm ISO 6983 [4]. The main principles of G-code programming date from the beginning of the Sixties, when machine tools were driven by punch cards. Since this time, huge improvements have been made on the technologies of machines and tools as well on the calculation capacities of the computer numerical control units. Nowadays, the G-code shows its limits and does not satisfy the requirements in term of programming any longer. Thus, manufacturers take more and more liberties to develop tools allowing filling the lacks of the G-code, each one proposing their own vendor-specific extensions of the original standard.

---
* Institut de Recherche en Communication et Cybernétique de Nantes, France.

Many works have led to huge progress allowing the birth and democratization of high speed machining, the increase of productivity with powerful machining cells. However, few works have been focused on the programming language itself. To fill this gap, a new data standard, STEP-NC, is under development and could deeply change our vision of the numerical chain CAD-CAM-CNC, by the implementation of a single file. Based on the Step standard for the exchange of products (ISO 10303), the STEP-NC file contains all the information for manufacture, through the description of machining entities, workingsteps, worplan, tools, machining strategies, etc. All these attributes are standardized in ISO 14649 norm [6] and are integrated in the corresponding application protocol ISO 10303-238 [5]. In the specific case of rough milling, the current standardized strategies are confined with classical ones as bidirectional, contour parallel, spiral, etc. New strategies built on the repetition of a pattern all along a guide curve can offer a lot of new possibilities and present serious advantages like their construction simplicity and fastness. Their integration in the STEP-NC standard is a further step in the development of these kinds of strategies, in addition to the enrichment of the possibilities offered by the STEP-NC standard. Two main aspects are developed in this paper: the generation of toolpaths using pattern strategies and the integration of these in STEP-NC standard. After a review of technologies concerning these two aspects, this paper will show a proposition of integration of pattern strategies in STEP-NC standard and its validation by the development of a complete STEP-NC numerical chain.

## 2. REVIEW OF STEP-NC TECHNOLOGY AND PATTERN STRATEGIES

### 2.1. STEP-NC STATE OF THE ART AND EXISTING PROTOTYPES

STEP-NC proposes a new vision of the numerical chain, and especially the place of manufacture in it. Many works were done to propose models in connection with this desire of total integration of the whole numerical chain [14, 15]. The basic principles are collected in the ISO 10303-238 application protocol [5] and in ISO 14649 norm [6], which is more machining oriented. STEP-NC file construction is based on decomposition of the part into machining entities, each one referring to geometry information and machining choices including machining strategies. All the workingsteps of the workplan can be ordered in different ways in the interpreter of the CNC, independently of the STEP-NC file [10]. In the case of a G-code file, the interpreter executes line by line every low level instructions of the program. In STEP-NC, the program is executed in terms of workingsteps. The interpretation is possible after a complete reading of the STEP-NC file in which some lines refer to others. This new way of proceeding leads to generate toolpaths at the shop floor, rather than in CAM. As a result, a part of the intelligence and power of decision is transferred to the computer numerical command. Self learning algorithms begin to be developed to produce better quality parts compensating controlled errors [8]. The interpreter has a central place at the manufacturing level and treats a large amount of tasks and calculations, which could affect and slow down the machining. However, Kramer et al. show that the method works for the criteria computation time [7]. Several interpreters and prototypes have already emerged. One of the first developed prototype is based on an industrial Siemens 840D CNC [13]. It was developed during the European ESPRIT project by the WZL laboratory in Aachen. Today, it doesn't seem to be under development any longer. A few other prototypes use plug-ins as for example ST-plan and ST-Machine, developed by Step TOOL Inc [3]. It allows using STEP-NC on a machine tool without the need to change its structure or command. Further works are lead in New Zealand by the team of Professor Xu allowing using STEP-NC on conventional G-code using machine [12].

An integrated STEP-NC numerical chain prototype has been developed by the team of Professor Suh in South Korea [11]. It is based on several modules, connected via a CORBA network. From the CAD file, the first module allows to generate STEP-NC file (PosSFP). Then, some other modules interpret the STEP-NC file, generate and allow visualisation of toolpaths (PosTPG and PosTPV). The numerical chain is completed by a human/machine interface and a module to communicate with machine tool axis cards. Some parts of the works presented in this paper were done on this Korean STEP-NC prototype, especially the implementation of the proposed model in the module PosSFP.

2.2. MACHINING STRATEGIES AVAILABLE IN CURRENT STEP-NC STANDARD

STEP-NC file includes all the machining information and the geometry of entities. These entities are intended to be machined and, in 2D milling, several strategies are proposed by the ISO 14649 norm (Fig. 1). STEP-NC file doesn't include explicit toolpaths as G code but only a high level description of the machining strategy. The 2D machining strategies defined in the norm are some of the most commonly used: bidirectional, parallel and spiral contour, and so on (Fig. 1). The attribute Explicit allows describing any other strategy by defining the whole of the toolpath. This attribute is used when the strategy cannot be described by one already normalized.

Thus, this range of strategies already allows a lot of possibilities. It is a good base for further work to lay on the principles for implementing strategies in STEP-NC standard.

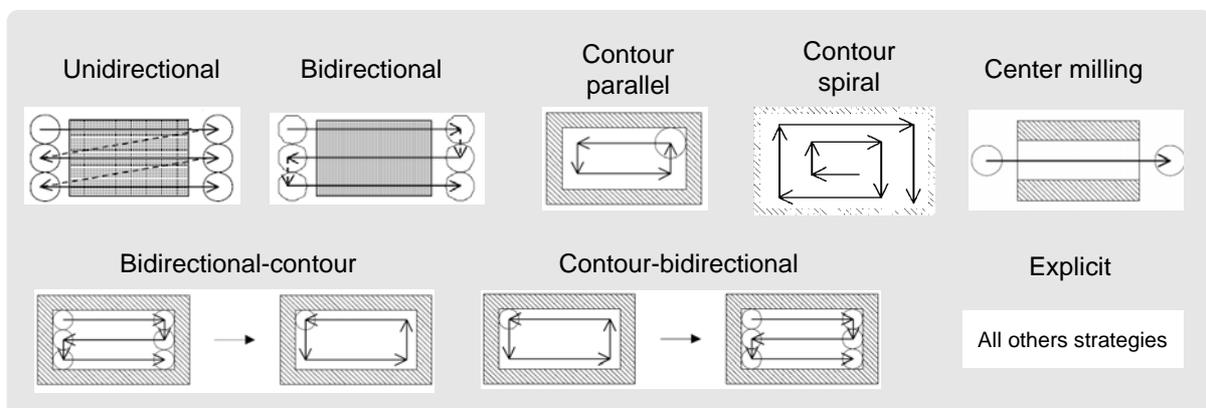

Fig. 1. Strategies proposed in STEP-NC norms

However, it does not fully take advantage of the opportunities offered by STEP-NC high-level information. Few developments have been performed in this area. There is a real interest to enrich the possibilities in this area and to make it more attractive. Pattern strategies bring a new approach for toolpath construction. In this paper, a proposition of the bases of their definition and implementation in STEP-NC standard is made.

### 2.3. PATTERN STRATEGIES

A pattern strategy can be defined as a strategy built on repetition of a pattern all along a guide curve. Some works are under development at the IRCCyN laboratory on this new way of building toolpaths [1]. Guide curve is built from the geometry of the machining entity contrary to the pattern, which is totally independent. The different applications can be milling strategies as trochoïdal milling or plunge milling, drilling, tapping, etc. There are several ways to give a mathematical definition of a pattern path. For the application treated in this paper (trochoïdal milling), a vectorial definition of the path can be used.

$$\overrightarrow{OM} = \vec{C}(t) + \overrightarrow{M}(t)$$

$\overrightarrow{OM}$ : Running point of toolpath.
$\vec{C}(t)$ : Guide curve built on geometric parameters of the entity.
$\overrightarrow{M}(t)$ : Pattern

This mathematical definition, based on a parametric description of the toolpath, needs some specific adaptations for each strategy. For some other applications, it is relevant to use function composition. In this paper, only a parametric definition of the toolpath is presented because it is well adapted in the exposed example: trochoïdal milling.

Indeed, trochoïdal milling toolpaths have been developed at the IRCCyN with this method for rough pocket milling [2]. The case of trochoïdal milling is particularly interesting because it is a new machining strategy, often unknown or ignored and which definition is not simple without using pattern strategies. Some CAM softwares build it as a succession of circles arcs (using G2 in G code format) linked with segments (G1). One

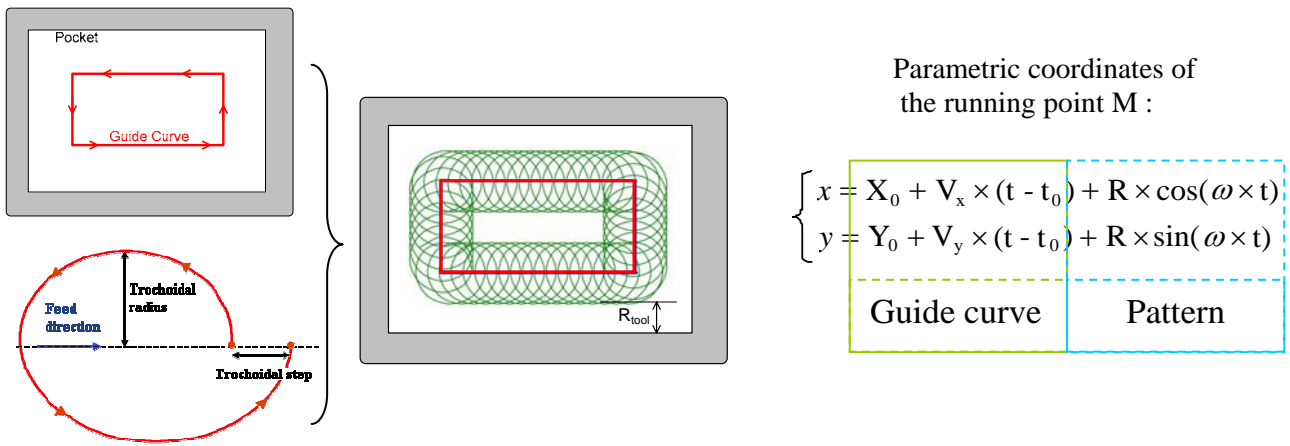

Parametric coordinates of the running point M :

$$\begin{cases} x = X_0 + V_x \times (t - t_0) + R \times \cos(\omega \times t) \\ y = Y_0 + V_y \times (t - t_0) + R \times \sin(\omega \times t) \end{cases}$$

Guide curve | Pattern

Fig. 2. Composition of two movements to build a trochoïdal toolpath

problem of this method comes from the acceleration discontinuity which is involved at the beginning of every circle arc. Some other methods of construction involve a large amount of points which can affect the G code file size and the computation capacities of the CNC. For the following applications, the trochoïdal toolpath is plane, for a constant Z level. The guide curve is, for this example, a succession of segments locally defined by a directing vector (Vx; Vy) and a starting point M0 (X0; Y0) corresponding to the value t0 of the parameter t (Fig. 2). The only necessary parameters to communicate in order to be able to build the toolpath are trochoïdal radius and trochoïdal step. The guide curve is built from the geometry of the machining entity. Thus, modification of the geometry of an entity only involves a modification of the guide curve, the pattern being independent. The contribution of such strategies for the construction of toolpaths is mainly in their definition simplicity, the possibilities of quick modifications and the formalism of definition which allows a lot of possible combinations. It is also possible to build pattern toolpaths automatically from a restricted number of parameters: pattern related parameters and the guide curve support strategy.

Simplicity of construction, fastness of the modifications and definition from a restricted number of parameters make pattern strategies a new way of programming toolpaths with a real interest and a particular compatibility with STEP-NC Standard.

## 3. INTEGRATION OF PATTERN STRATEGIES IN STEP-NC NUMERICAL CHAIN

The proposal to implement pattern strategies in STEP-NC standard rests on already normalized strategies for the construction of guide curve. Then, this proposition has been validated by creating a complete STEP-NC numerical chain. This allows the simulation, in the case of trochoïdal milling, of the use of pattern strategies from the CAD file to the real machined part. The pattern strategies have been implemented here in the Two5DmillingStrategy function [5], represented here in Express G formalism (Fig. 3). This function deals with the definition of parameters and attributes of machining strategies. To build the pattern, several kinds of attributes can be chosen: length measure, direction, etc. In

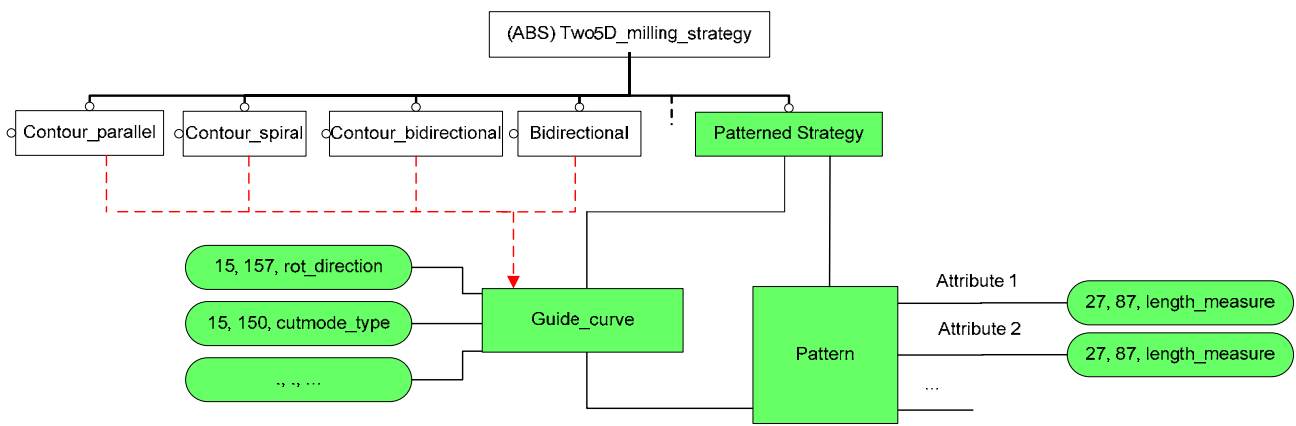

Fig. 3. ARM diagram completed with pattern strategies

the case of trochoïdal milling, the pattern parameters are trochoïdal radius and trochoïdal step. From these two attributes, the interpreter is able to build the pattern. This pattern is repeated all along the guide curve which gives the machining direction of propagation. The construction of the guide curve is made from geometric data of the machining entity, chosen support strategy, pattern parameters and characteristics of the tool. All these information must be written in the STEP-NC file, and the following syntax is adopted:

***#Ref=STRATEGY(overlap,allow_multiple_passes,#guide_curve,pattern parameters)***

Overlap and allow_multiple_passes are two common parameters for all strategies. Then, we find the definition of guide curve and pattern parameters. The definition of guide curve support strategy (#guide_curve) can be transferred to another line for more clarity. In a first time, the guide curve support strategy should be described by the existing strategies in STEP-NC norm. So guide curve can be built as bidirectional, parallel contour, spiral contour, etc. As a consequence, a range of trochoïdal strategies can be proposed by the simple modification of this attribute. This model was implemented in the Korean PosSFP software which allows generating STEP-NC file from CAD file. This extension of PosSFP was realized in two particular cases of pattern strategies: trochoïdal milling and plunge milling (Fig. 4). In this paper, only the case of trochoïdal milling is treated. So for the example of trochoïdal milling, all the parameters needed for the construction of the toolpath are grouped in the STEP-NC file with the following syntax:

*TROCHOIDAL(overlap,allow_multiple_passes,guide curve support strategy,parameters of guide curve, trochoïdal radius, trochoïdal step)*

By reading these attributes, the geometry of the machining entity and the characteristics of the tool, the interpreter is able to build the whole toolpath. The implementation of pattern strategies in the STEP-NC file is a first step. Its validation in a STEP-NC numerical chain environment is important to validate the concept. The following steps are intended to fulfill this objective.

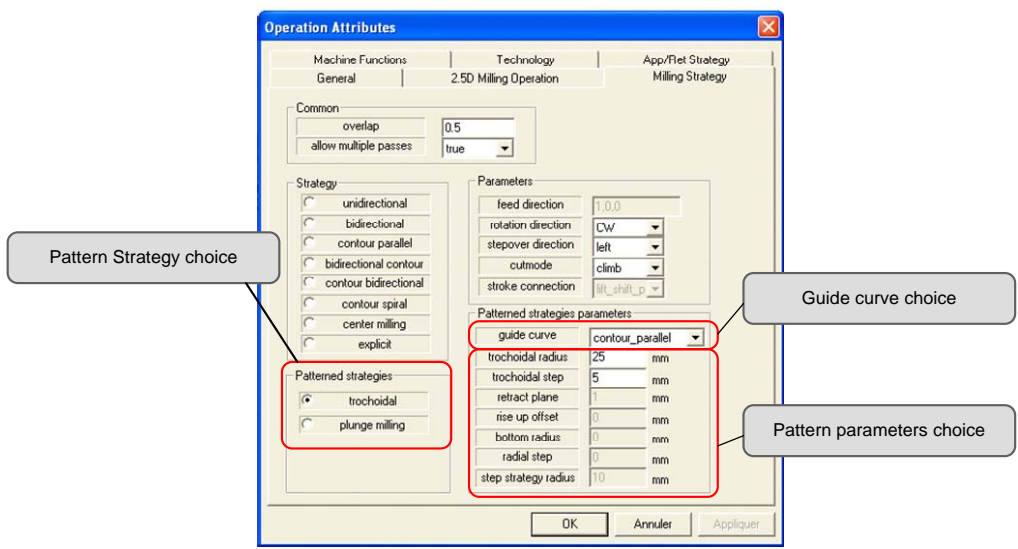

Fig. 4. Implementation of pattern strategies developed in the PosSFP Korean software

## 4. THE COMPLETE STEP-NC NUMERICAL CHAIN DEVELOPED AT IRCCYN

In order to validate the implementation of pattern strategies in the STEP-NC standard, a complete STEP-NC numerical chain was created. Therefore, we can test both the generation of the STEP-NC program and its interpretation until the machining of test parts in the case of the trochoïdal milling. The STEP-NC numerical chain relies on three main modules (Fig. 5):

- The first module is mainly made of the Korean software PosSFP. It has been doted with pattern strategies and can generate the STEP-NC file from a CAD file. It is based on recognition of the machining entities. Moreover, it makes possible the selection of all machining parameters which will be used afterwards in the STEP-NC file. The implementation of pattern strategies in the Korean software PosSFP was made from the source code of PosSFP kindly lent by Professor Suh's team from the University of Postech in South Korea.

- Thanks to the second module that was developed at IRCCyN, it is possible to read the STEP-NC file generated at the previous step. It also allows the generation and display of the toolpaths. From entity geometrical data, characteristics of the tools and the pattern strategy parameters, the interpreter can create a toolpath. It builds successively the guide curve and repeats the pattern all along it. A visualization module allows user to see generated toolpath and validate it. The output is a file that contains control points.

- The third module processes the file in order to read and to execute toolpaths thanks to our production mean (Fatronik Verne parallel kinematics machining center [9] equipped with a CNC Siemens 840D).

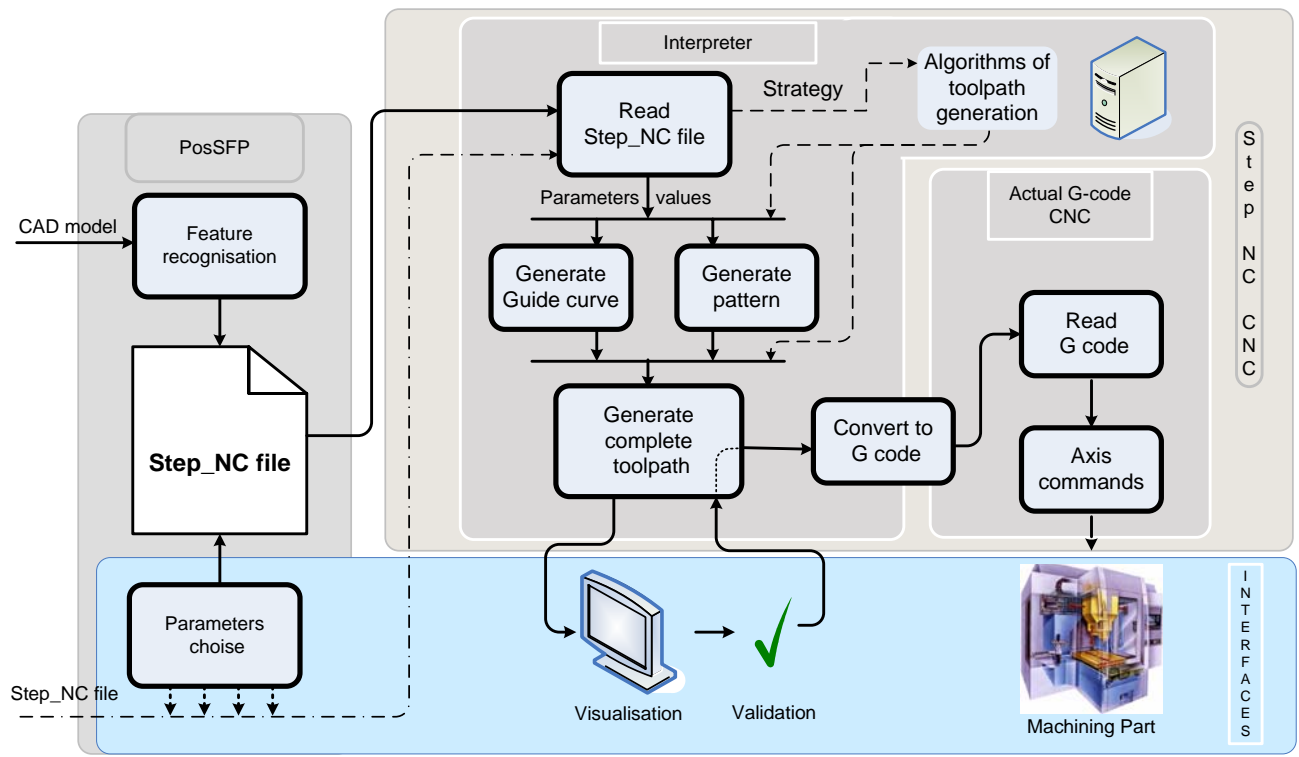

Fig. 5. STEP-NC numerical chain to generate pattern toolpaths

To machine the part on current production means, toolpath must be adapted to G code. This is the object of the third module. In a totally STEP-NC integrated CNC, the toolpath could be directly processed on machine tool axis. In the example of a rough pocket milling (Fig. 6), three STEP-NC files were generated: the first two use a guide curve based on parallel contour strategy (1 and 2, Fig. 6) but have different pattern parameters and trochoïdal radius. The first trochoïdal radius permits to mill the whole pocket with only one trochoïdal parallel contour. When the whole pocket is not machined by only one trochoïdal parallel contour, there are two cases. Either it is interesting to do another trochoïdal contour on the remaining material or only a median trochoïdal milling is necessary (case 2). So, it is easy to play on the guide curve support strategy. For example, the third case (3, Fig. 6), has a birectional guide curve. The only thing that changes in the STEP-NC file is the attribute #guide_curve. That means changing only one attribute can change all the toolpath. If this modification had to be done with a G code file, the whole file would have been changed.

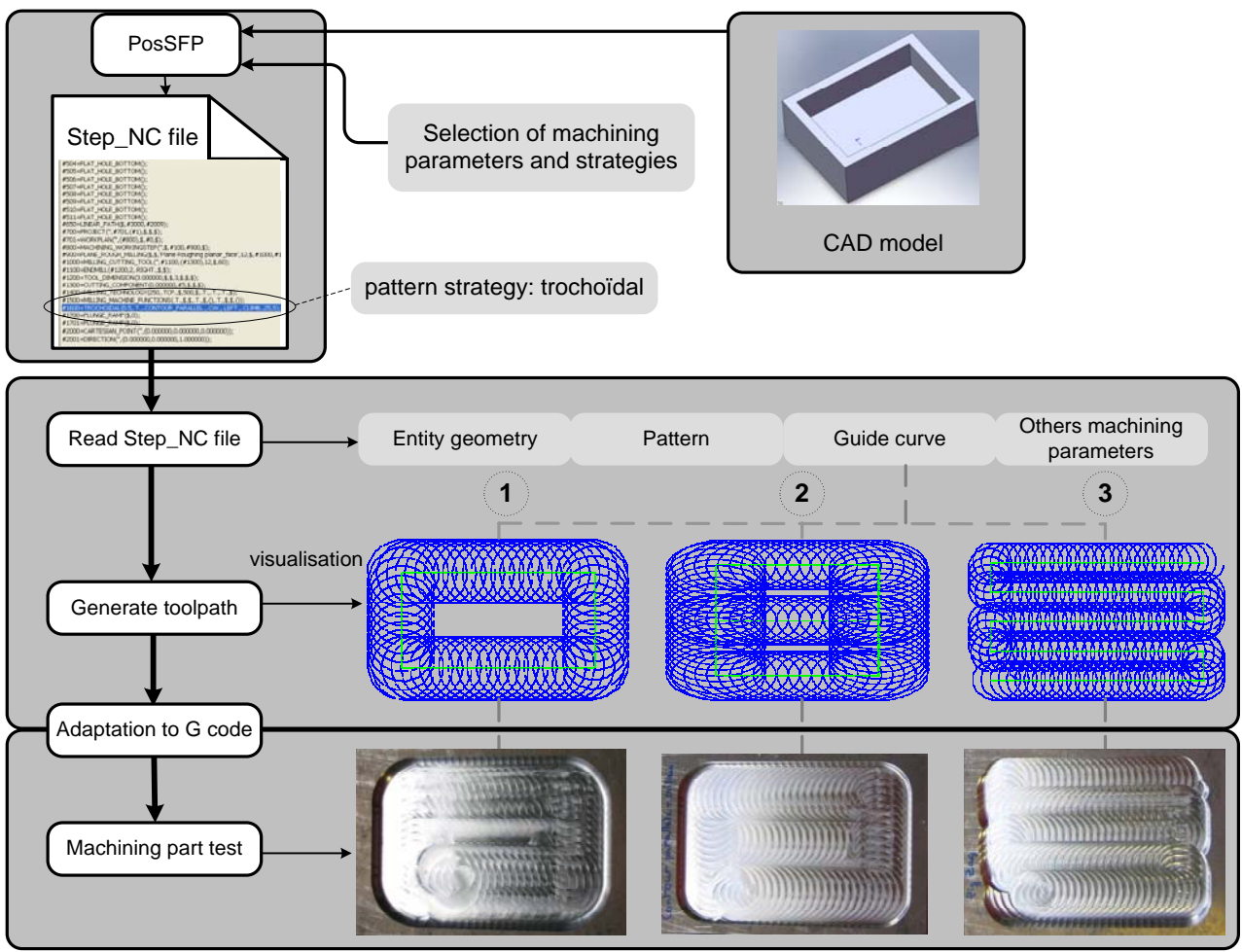

Fig. 6. Trochoïdal toolpaths built on pattern strategy in STEP-NC numerical

## 5. RESULTS, DISCUSSIONS AND FURTHER DEVELOPMENTS

The developed interpreter can generate automatically all the operations from reading the STEP-NC file to the machining part. Only the adaptation to G-code that allows production on our means still requires external intervention but the goal is to remove it ultimately. The choice for the definition of pattern strategies was tested to show the interest of their construction properties (simplicity, fastness, etc.) and the resulting possibilities for the automation of toolpath generation. Several parts were machined with different pocket geometries, guide curves and patterns. This allowed validating the adaptability of the developed tools by assessing their response to entity or technological parameters modifications. The results allowed verifying the validity of the approach of toolpaths construction based on pattern strategies and their implementation in STEP-NC standard.

So, STEP-NC offers a lot of possibilities to put forward the advantages of pattern strategies and, in general, of new machining strategies that G code bridles. Conversely, the introduction of these innovative strategies brings STEP-NC legitimacy compared to G code. Indeed, it does not restrict it to the use of common strategies but shows its strength through strategies that are more sophisticated and often misused because of the difficulty to program them in G code. This allowed either to show that it is possible to create concrete solutions to participate to the development of STEP-NC standard without owning especially a complete prototype of STEP-NC numerical chain. In a first part, common production means (programmed with G-code) can be used by interpreting the STEP-NC file. This way to do is enough for some applications, as in the case of implementation of new strategies in STEP-NC. However, it can quickly become a major handicap insofar as the internal feed-back in the CNC is made impossible by the use of G code. Some more advanced applications, such as real-time toolpath correction or machining events management, will face this problem. Using a completely integrated STEP-NC CNC will be an inevitable stage. STEP-NC is not intended to equalize G code but to exceed it by proposing new and innovating possibilities. Thus, we are now in a first phase of development and experimentation starting from interpreted STEP-NC programming. This phase requires an adaptation to our production means by interpreting to G code (Fig. 7). The next stages must first of all enable us to be totally freed from the G code by directly piloting machine axis from the STEP-NC interpreter (Fig. 9). This step will allow integrated STEP-NC programming and will open the way for more concrete developments of this standard. From this, the advanced STEP-

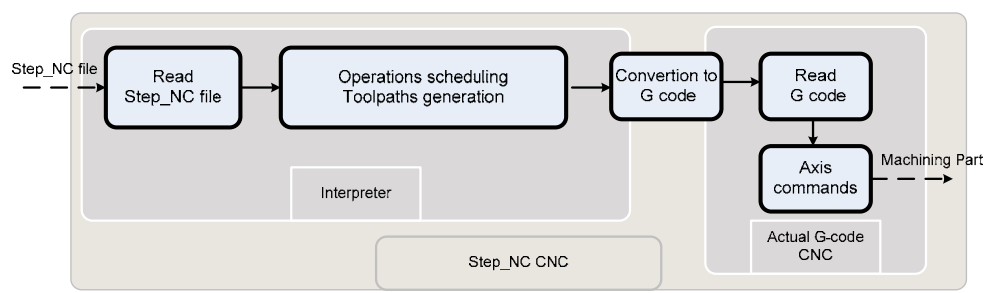

Fig. 7. Interpreted STEP-NC programming

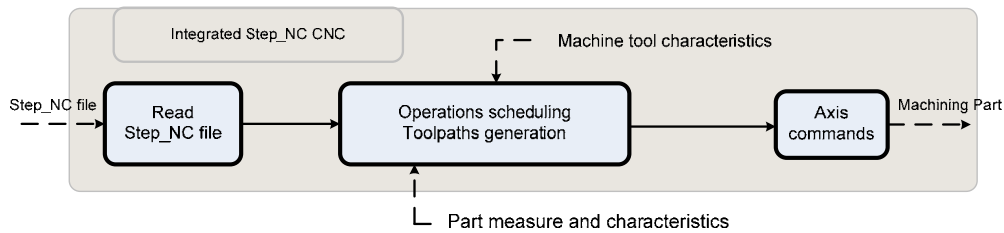

Fig. 9. Integrated STEP-NC programming

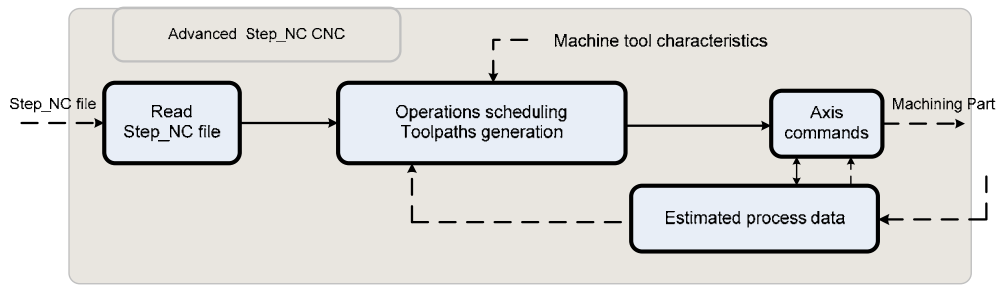

Fig. 8. Advanced STEP-NC programming

NC programming will constitute the ultimate stage. It includes estimating process data during machining (Fig. 8). This will permit real time adaptation of machining parameters of as well as trajectories. The autonomy of the CNC will thus be increased for additional optimizations of the machining workplan. The other concept exposed in this paper is the construction of toolpaths using pattern strategies. This method can be very interesting in other cases than trochoïdal milling: a pattern strategy definition is particularly adapted for plunge milling by the repetition of a plunge pattern all along a guide curve. Some other manufacturing proceeds can also benefit from this construction method, such as applications in incremental sheet forming. Moreover, the combined use of several strategies for machining a same entity can have advantages and can be easily set up in STEP-NC. It is already the case for the association parallel-bidirectional contour in STEP-NC norm. Trochoïdal milling can be very efficient in the case of opening a pocket. It can then be finished with another strategy. Moreover, pattern parameters could be adapted according to the pocket geometry and the capacities of the machine tool and be adapted for a same entity.

All these prospects and further possible developments take a particular interest with the use of STEP-NC standard. It opens a new range of possibilities, and particularly in the research of innovating toolpaths that are more adapted to new possibilities of machine tools.

ACKNOWLEDGEMENTS

The authors really wish to thank professors Suh from the University of Postech and his research team to have had the kindness and confidence to give us at the disposal of some of their development tools.